\documentclass[lettersize,journal]{IEEEtran}
\IEEEoverridecommandlockouts
\usepackage{amsmath,amsfonts}
\usepackage{algorithmic}
\usepackage[ruled,linesnumbered]{algorithm2e}
\usepackage{array}
\usepackage[caption=false,font=normalsize,labelfont=sf,textfont=sf]{subfig}
\usepackage{textcomp}
\usepackage{booktabs}
\usepackage{multirow}
\usepackage{booktabs}
\usepackage{stfloats}
\usepackage{url}
\usepackage{verbatim}
\usepackage{graphicx}
\usepackage{cite}
\usepackage{authblk}
\usepackage{hyperref} 
\hypersetup{
    colorlinks=true, 
    citecolor=blue,
    linkcolor=blue,   
    filecolor=blue,     
    urlcolor=black       
}

\title{Stealth Signals: Multi-Discriminator GANs for Covert Communications Against Diverse Wardens}

\author{
\IEEEauthorblockN{Afan Ali$^1$, Md. Jalil Piran$^2$,~\IEEEmembership{Senior Member,~IEEE,} and H\"{u}seyin Arslan$^1$,~\IEEEmembership{Fellow,~IEEE}}\\
\IEEEauthorblockA{$^1$School of Engineering and Natural Sciences, Istanbul Medipol University, Istanbul, Türkiye\\
Email: \{afanali85@gmail.com, huseyinarslan@medipol.edu.tr\}}\\
\IEEEauthorblockA{$^2$Department of Computer Science and Engineering, Sejong University, Seoul, South Korea\\
Email: piran@sejong.ac.kr}
}

\begin{document}

\maketitle

\begin{abstract}
Covert wireless communications are critical for concealing the existence of any transmission from adversarial wardens, particularly in complex environments with multiple heterogeneous detectors. This paper proposes a novel adversarial AI framework leveraging a multi-discriminator Generative Adversarial Network (GAN) to design signals that evade detection by diverse wardens, while ensuring reliable decoding by the intended receiver. The transmitter is modeled as a generator that produces noise-like signals, while every warden is modeled as an individual discriminator, suggesting varied channel conditions and detection techniques. Unlike traditional methods like spread spectrum or single-discriminator GANs, our approach addresses multi-warden scenarios with moving receiver and wardens, which enhances robustness in urban surveillance, military operations, and 6G networks. Performance evaluation shows encouraging results with improved detection probabilities and bit error rates (BERs), in up to five warden cases, compared to noise injection and single-discriminator baselines. The scalability and flexibility of the system make it a potential candidate for future wireless secure systems, and potential future directions include real-time optimization and synergy with 6G technologies such as intelligent reflecting surfaces.
\end{abstract}


\section{Introduction}
\IEEEPARstart{T}{he} primary objective of private and covert communications is to conceal the existence of wireless data transmission from adversarial entities, often referred to as wardens, thereby providing a robust layer of security. There are various approaches for covert communications, such as spreading signals below noise level, embedding them into ambient or artificial noise, or leveraging inherent ambiguities such as channel variations and location uncertainty, which significantly weaken the detection capabilities of potential adversaries. The advent of wireless networks, particularly with the emergence of 5G and ongoing development of 6G, has enabled access to progressively complex adversarial environments. These environments involve various wardens employing diverse detection mechanisms, such as machine learning-based detectors and sophisticated signal processing detectors, and necessitate new methods for secure and hidden communication. Spread spectrum technique is one of the prominent method to conceal communication from unintended users, which spreads the signal over a wide bandwidth so that its power spectral density (PSD) is reduced, making it appear noise-like and not detectable by the enemy. In ~\cite{24,25}, authors introduced direct-sequence spread spectrum (DSSS) and frequency-hopping spread spectrum (FHSS) to conceal transmission presence through signal blending with noise or rapidly hopping frequencies. 

Recent developments in artificial intelligence (AI) have enabled new modes of enhancing clandestine communications. Techniques based on AI, such as, deep neural networks (DNNs), have been investigated for generating signals that learn adaptively how to utilize channel uncertainties to evade detection.  Authors in \cite{1} provide a comprehensive survey of covert communication methods, such as, power control and noise injection for concealing from detection. Similarly, work in \cite{2,3,4,14}  investigate the employment of neural networks in adversarial environments to craft covert signals. Additionally, artificial noise injection has been used to conceal transmissions by adding controlled interference, which also confuses the warden in distinguishing authentic signals from ambient noise~\cite{1,3,5,6}. These traditional methods, though effective in single-warden systems, are inept while handling heterogeneity and multiplicity of wardens in modern wireless environments, such as urban surveillance networks or military operations. 
Federated learning (FL) has also emerged as a promising approach, allowing distributed devices to collaborate for training models for covert signal design without sensitive data sharing, hence ensuring privacy while maintaining security~\cite{22,23}. FL is specifically preferable in situations where devices operate in decentralized scenarios in order to prevent random eavesdropping~\cite{22}. Table~\ref{tab:ai_covert} summarizes key works leveraging AI for covert and private communications. 

Generative AI (GAI) based methods, such as Generative Adversarial Networks (GAN), have also drawn attention to generate noise-like realistic signals that can deceive adversaries~\cite{5,26,27}. It  trains a generator to generate signals that can mimic noise, deceiving a discriminator that is imitating an adversary's detection process. However, most of these GAN approaches make the assumption of a single warden or homogeneous detection abilities, limiting their capabilities in multi-warden scenarios where adversaries have different channel conditions, noise patterns, or detection algorithms~\cite{26}. These methods can suffer from training instability and do not generalize between disparate wardens, leading to elevated detection probabilities under realistic conditions~\cite{28}. In this work, we investigate multi-discriminator GAN to overcome these deficits by modeling each warden as a single discriminator, allowing the generator to learn a signal design that avoids all wardens simultaneously and yet provides reliable decoding by the intended receiver. This approach will improve robustness and scalability, attaining lower detection probabilities and bit error rates (BERs) compared to single-discriminator GANs and traditional noise injection methods, for urban surveillance, military and 6G applications.
 
\begin{table*}[t]
\caption{Summary of Works Using AI for Covert and Private Communications}
\label{tab:ai_covert}
\centering
\begin{tabular}{p{4cm} p{2cm} p{5cm} p{2cm} p{2cm}}
\toprule
 \textbf{AI Technique} & \textbf{No. of wardens} &  \textbf{Application/Focus} &   \textbf{Reference} & \textbf{Year}  \\
\midrule
 Autoencoders & 1 & Autoencoders based  Signal encoding for covertness & \cite{2,3} & 2023 \\
Deep learning (DL) & 1 & DL based Random eavesdropping mitigation & \cite{5,31} & 2024 \\
Neural Networks (NN)  & 1& Using NN in Privacy for 6G networks & \cite{13,32} & 2022 \\
 DL-Based Channel Prediction & 1 & Deep Neural Networks (DNN) for Channel uncertainty exploitation & \cite{14} & 2020 \\
 Federated Learning (FL) & 1  & Using FL for Private data aggregation & \cite{15} & 2019 \\
Convolutional Neural Networks (CNN) & 1 & CNN based Noise pattern generation  & \cite{14,33} & 2023 \\
 Differential Privacy in Comm. & 1 & Differential Privacy for Signal indistinguishability & \cite{16,33} & 2024 \\
 Beamforming for Covertness & 1 & RL in Directional covert transmission & \cite{17,34} & 2024 \\
 AI Noise Swarms & 1 & Multi-Agent RL in Dynamic interference fields  & \cite{18,35} & 2023 \\
Quantum-Inspired AI Covert & 1 & DNN in Fragmented signal design & \cite{19} & 2022 \\
\bottomrule
\end{tabular}
\end{table*}

\subsection{Motivation and Contribution}
The motivation for this research arises from growing complexity in adversarial environments, where there are multiple wardens with different detection abilities, such as, different antenna configurations, locations, and algorithms, which pose significant challenges to covert communication systems. Traditional methods are unable to deal with such diversity, and they tend to raise detection risk under practical applications~\cite{7}. To the best of our knowledge, this idea has not yet been reported by the literature. The main contributions of our work are summarized as follows.

\begin{itemize}
  \item  We propose a generative AI (GAI) based framework to facilitate the covert communication against multi-wardens threat in a communication system. Although most methods for covert communication consider single-warden and static scenario, we are the first to introduce a GAI based signal design to handle the existence of different warden detection capabilities amidst moving wardens and receiver, while making it possible for the intended receiver to receive the signal.
  \item Under the proposed GAI method, we design a novel end-to-end adversarial GAN system for generating stealthy signals, which is the cornerstone of the transmitter's operation. Transmitter, modeled as a generator, produces complex-valued signal, $\mathbf{s}$, that encodes a hidden message, random noise and is shaped to compensate expected channel delays. The random noise vector introduces variability to make $\mathbf{s}$ appear as environmental noise to wardens, enhancing covertness. 
  \item Similarly, we design each warden as a discriminator that attempts to classify noise-only observations from transmission observations. Heterogeneity in channel model and noise variance of each warden ensures each discriminator learns an independent detection strategy, which depicts real-world warden heterogeneity.
  \item Consequently, this adversarial game continues iteratively, with generator and discriminators updating their parameters to outmaneuver each other, converging to a solution where \( \mathbf{s} \) is covert to all wardens yet decodable by Bob. We provide a clear implementation framework, training methods, explaining how it achieves covertness objective and overall optimization in the adversarial game, and testing standards to ensure realistic usability.
  \item We provide adequate performance results to validate the effectiveness of our proposed method to work in a covert communication. In particular, we establish the usability of this approach in dynamic, multi-adversary scenarios and adaptive wardens with AI algorithms, giving an enormous advancement in stealthy and private communication security.
\end{itemize}

The remainder of this paper is organized as follows: Section~\ref{sec:system_model} presents the system model and problem formulation. Section~\ref{sec:proposed_framework} details the proposed adversarial AI framework, including the multi-discriminator GAN architecture, training methodology and derivation for loss functions. Section~\ref{sec:simulation} discusses simulation results and performance evaluation, while Section~\ref{tab:conclusion} concludes the paper with future research directions.

\section{System Model and Problem Formulation}
\label{sec:system_model}

\begin{figure}[t]
    \centering
    \includegraphics[width=0.7\columnwidth]{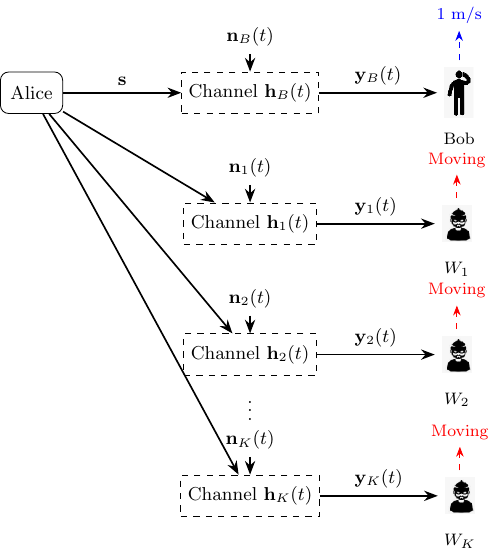}
    \caption{System Model: Alice sends $\mathbf{s}$ to moving Bob over a shared channel. Moving wardens $W_i$ attempt to detect $\mathbf{s}$ using binary hypothesis testing.}
    \label{fig:sys_model}
\end{figure}

\begin{table}[!t]
\centering
\small
\caption{List of frequently Used Notations}
\label{tab:notations}
\begin{tabular}{@{}ll@{}}
\toprule
\textbf{Notation} & \textbf{Description} \\
\midrule
Alice & Transmitter \\
Bob & Receiver \\
$W_i$ & $i$-th warden  \\
$K$ & Number of wardens \\
$m $ & Hidden message  \\
$\mathbf{s} $ & Transmitted signal \\
$N$ & Signal length (samples per time slot) \\
$P$ & Maximum transmit power \\
$\mathbf{z} $ & Random Gaussian noise vector \\
$\mathbf{h}_B $ & Channel vector from Alice to Bob \\
$\mathbf{h}_i $ & Channel vector from Alice to $W_i$ \\
$\mathbf{y}_B$ & Received signal at Bob with noise \\
$\mathbf{x}_B$ & Received signal at Bob without noise \\
$\mathbf{y}_i$ & Received signal at $W_i$ \\
$\mathbf{n}_B $ & Noise at Bob \\
$\mathbf{n}_i $ & Noise at $W_i$ \\
$\sigma_B^2$ & Noise variance at Bob \\
$\sigma_i^2$ & Noise variance at $W_i$ \\
$\mathcal{G}$ & Generator (transmitter, Alice) \\
$\mathcal{D}_i$ & Discriminator for warden $W_i$ \\
$\mathcal{D}_{\text{Bob}}$ & Decoder at Bob \\
$\mathcal{L}_i$ & Loss function for discriminator \\
$\mathcal{L}_G$ & Loss function for the generator \\
$\lambda_i$ & Weight for $W_i$’s contribution to generator loss \\
$\mu$ & Weight for decoding reliability in generator loss \\
$\mathcal{L}_{\text{decode}}$ & Decoding loss \\
$\hat{m}$ & Decoded message at Bob \\
$H_0$ & Hypothesis: No transmission \\
$H_1$ & Hypothesis: Transmission \\
$P_{D,i}$ & Detection probability at $W_i$ \\
$P_{F,i}$ & False alarm probability at $W_i$ \\
$\epsilon$ & Covertness threshold for $P_{D,i}$ \\
$P_e$ & Error probability at Bob \\
$D_{\text{KL}}$ & Kullback-Leibler divergence between $H_0$ and $H_1$ \\
$\delta$ & Covertness threshold for $D_{\text{KL}}$ \\
\text{BER} & Bit Error Rate at Bob \\
\text{CSR} & Covertness Success Rate \\
$\mathbf{R}_i$ & Spatial correlation matrix for channel \\
$\rho$ & Correlation coefficient \\
$f_d$ & Doppler shift \\
$\tau_l$ & Delay of the $l$-th tap \\
$T_s$  & Symbol duration \\
\bottomrule
\end{tabular}
\end{table}

\subsection{Communication Scenario}
\label{sys_model}
Figure~\ref{fig:sys_model} illustrates a detail block diagram showing communication scenario. We consider a wireless communication system that consists of a transmitter (Alice), a desired receiver (Bob), and a set of \( K \) adversarial wardens (Willies), denoted as \( \{W_1, W_2, \ldots, W_K\} \). Alice would like to send a hidden message \( m \in \mathcal{M} \) to Bob over a shared wireless channel in a way that no warden can reliably determine the presence of the transmission. The wardens are passive players that possess detection algorithms capable of distinguishing the presence or absence of a transmission based on what they observe from the received signal. Table~\ref{tab:notations} lists the notations frequently used in the derivations later.

Alice's signal is denoted as \( \mathbf{s} \in \mathbb{C}^N \), with \( N \) being the signal length denoted by sample number of a time slot. The signal is also subject to a power constraint \( \| \mathbf{s}\|^2 \leq P \), with \( P \) being the maximum transmit power. To model realistic channel effects, we incorporate spatial correlation, mobility, and frequency diversity into the channel models for Bob and the wardens. The channel between Alice and Bob, $\mathbf{h}_B(t) \in \mathbb{C}^N$, is time-varying and frequency-selective due to Bob's mobility and multi-path propagation. It is modeled as a frequency-selective channel as:
\begin{equation}
\mathbf{h}_B(t) = \sum_{l=0}^{L-1} h_{B,l}(t) \delta(t - \tau_l),
\label{bob_channel}
\end{equation}
where $h_{B,l}(t)$ is the complex gain of the $l$-th tap at time $t$ and $\tau_l = l T_s$ is the delay of the $l$-th tap with $T_s$ being the symbol duration. The tap gains $h_{B,l}(t)$ exhibit spatial correlation and Doppler effects due to mobility. The spatial correlation is modeled by a correlation matrix $\mathbf{R}_B$, where $[\mathbf{R}_B]_{m,n} = \rho^{|m-n|}$ where $\rho$ is the correlation coefficient, reflecting correlation across the $N$ samples within a time slot. The time variation due to mobility is modeled with a Doppler shift $f_d$, corresponding to Bob moving at a walking speed of 1 m/s. Thus, $h_{B,l}(t)$ evolves as:
\begin{equation}
h_{B,l}(t) = \tilde{h}_{B,l} e^{j 2 \pi f_d t},
\end{equation}
where $\tilde{h}_{B,l} \sim \mathcal{C N}(0, \mathbf{R}_B)$ incorporates the spatial correlation. The received signal at Bob, accounting for the frequency-selective channel, is given by the convolution of $\mathbf{s}$ with $\mathbf{h}_B(t)$:

The received signal at Bob is:
\begin{equation}
\mathbf{y}_B(t) = \sum_{l=0}^{L-1} h_{B,l}(t) \mathbf{s}(t - \tau_l) + \mathbf{n}_B(t),
\label{Rx_signalB}
\end{equation}
where $\mathbf{n}_B(t) \sim \mathcal{C N}(0, \sigma_B^2 \mathbf{I})$ is Bob's additive white Gaussian noise (AWGN) with variance $\sigma_B^2$. Bob estimates the channel $\mathbf{h}_B(t)$ using pilot symbols, resulting in an estimate $\hat{\mathbf{h}}_B(t) = \mathbf{h}_B(t) + \mathbf{e}_B$, where $\mathbf{e}_B \sim \mathcal{C N}(0, \sigma_e^2 \mathbf{I})$ models estimation errors with variance $\sigma_e^2$. For simplicity, we assume that Bob has a perfect channel state information (CSI), hence $\sigma_e^2=0$ and $\hat{\mathbf{h}}_B(t)=\mathbf{h}_B(t)$. Here we can define effective signal $\mathbf{x}_B(t)$, after removing noise, at Bob as:
\begin{equation}
\mathbf{x}_B(t) = \sum_{l=0}^{L-1} h_{B,l}(t) \mathbf{s}(t - \tau_l)
\label{effective_signalBob}
\end{equation}
For the $i$-th warden $W_i$, the channel from Alice, $\mathbf{h}_i(t) \in \mathbb{C}^N$, is similarly modeled as a frequency-selective, time-varying channel:
\begin{equation}
\mathbf{h}_i(t) = \sum_{l=0}^{L-1} h_{i,l}(t) \delta(t - \tau_l),
\label{warden_channel}
\end{equation}
with $h_{i,l}(t) = \tilde{h}_{i,l} e^{j 2 \pi f_d t}$, where $\tilde{h}_{i,l} \sim \mathcal{C N}(0, \mathbf{R}_i)$, and $\mathbf{R}_i$ is the correlation matrix with $[\mathbf{R}_i]_{m,n} = \rho^{|m-n|}$. The $f_d$ assumes the wardens may also experience relative motion. The received signal at $W_i$ is:
\begin{equation}
\mathbf{y}_i(t) = \sum_{l=0}^{L-1} h_{i,l}(t) \mathbf{s}(t - \tau_l) + \mathbf{n}_i(t),
\label{Rx_signalI}
\end{equation}
where $\mathbf{n}_i(t) \sim \mathcal{C N}(0, \sigma_i^2 \mathbf{I})$ is the AWGN at $W_i$ with variance $\sigma_i^2$. The channel $\mathbf{h}_i(t)$ differs across wardens due to their distinct locations, antenna configurations, or environmental factors, and is assumed to be unknown to Alice but partially characterizable through statistical models or historical data.

Each warden $W_i$ employs a detection mechanism, modeled as a binary hypothesis test:
\begin{itemize}
    \item $H_0$: No transmission occurs, and \eqref{Rx_signalI} simplifies to:
    \begin{equation}
    \mathbf{y}_i(t) = \mathbf{n}_i(t),
    \label{H0}
    \end{equation}
    \item $H_1$: Transmission occurs, resulting in \eqref{Rx_signalI}.
\end{itemize}
The warden's detection performance is quantified by the probability of detection:
\begin{equation}
P_{D,i} = \operatorname{Pr}(\text{decide } H_1 \mid H_1),
\end{equation}
and the probability of false alarm:
\begin{equation}
P_{F,i} = \operatorname{Pr}(\text{decide } H_1 \mid H_0).
\end{equation}

To derive $P_{D,i}$, assume each warden uses a likelihood ratio test (LRT). Under $H_0$, the received signal $\mathbf{y}_i(t)$ is given by \eqref{H0}, with probability density function (PDF):
\begin{equation}
p(\mathbf{y}_i(t) \mid H_0) = \frac{1}{(\pi \sigma_i^2)^N} \exp\left(-\frac{\|\mathbf{y}_i(t)\|^2}{\sigma_i^2}\right).
\end{equation}
Under $H_1$, the received signal is given by \eqref{Rx_signalI}. The PDF is:
\begin{equation}
p(\mathbf{y}_i(t) \mid H_1) = \frac{1}{(\pi \sigma_i^2)^N} \exp\left(-\frac{\|\mathbf{y}_i(t) - \mathbf{x}_i(t)\|^2}{\sigma_i^2}\right).
\end{equation}
The likelihood ratio ($\Lambda$) is written as:

\begin{align}
&\Lambda(\mathbf{y}_i(t)) = \frac{p(\mathbf{y}_i(t) \mid H_1)}{p(\mathbf{y}_i(t) \mid H_0)} \\ \notag
&= \exp\left(\frac{2 \operatorname{Re}(\mathbf{y}_i(t)^H \mathbf{x}_i(t)) - \|\mathbf{x}_i(t)\|^2}{\sigma_i^2}\right),
\label{eq:lrt}
\end{align}
where:
\begin{equation}
\|\mathbf{x}_i(t)\|^2 \approx \sum_{l=0}^{L-1} |h_{i,l}(t)|^2 \|\mathbf{s}\|^2,
\end{equation}
and $|h_{i,l}(t)|^2 = |\tilde{h}_{i,l}|^2$ since the Doppler phase $e^{j 2 \pi f_d t}$ has magnitude 1. The spatial correlation $\mathbf{R}_i$ affects the distribution of $\tilde{h}_{i,l}$. The warden decides $H_1$ if $\Lambda(\mathbf{y}_i(t)) > \gamma$, where $\gamma$ is a threshold set to satisfy $P_{F,i} \leq \alpha$. The test statistic $T = 2 \operatorname{Re}(\mathbf{y}_i(t)^H \mathbf{x}_i(t))$ is Gaussian under $H_1$ with mean $2 \|\mathbf{x}_i(t)\|^2$ and variance $2 \sigma_i^2 \|\mathbf{x}_i(t)\|^2$.
Thus:
\begin{equation}
P_{D,i} = Q\left(\frac{\sigma_i^2 \ln \gamma - \|\mathbf{x}_i\|^2}{\sqrt{2 \sigma_i^2 \|\mathbf{x}_i\|^2}}\right),
\label{eq:pd_derive}
\end{equation}
where $Q(\cdot)$ is the Q-function. The spatial correlation $\mathbf{R}_i$ and Doppler shift $f_d$ affect the distribution of $|h_{i,l}(t)|^2$. To decode the $m$ embedded in $\mathbf{s}$, Bob uses a secret key shared with Alice, which allows him to interpret the signal while the wardens attempt the presence of the transmission based solely on statistical differences between $H_1$ and $H_0$. Hence, the goal of covert communication is to ensure that \( P_{D,i} \) in \eqref{eq:pd_derive} remains below a threshold \( \epsilon \) for all \( i \), making the transmission indistinguishable from noise to all wardens.

 To formalize covertness, we use the Kullback-Leibler (KL) divergence between the distributions under \( H_0 \) and \( H_1 \):

\begin{equation}
D_{\mathrm{KL}}(p(\mathbf{y}_i \mid H_1) \| p(\mathbf{y}_i \mid H_0)) \approx \sum_{l=0}^{L-1} \frac{\mathbb{E}_t[|h_{i,l}|^2] \|\mathbf{s}\|^2}{\sigma_i^2},
\label{eq:kl}
\end{equation}
where $\mathbb{E}_t[|h_{i,l}(t)|^2] = \mathbb{E}[|\tilde{h}_{i,l}|^2]$ due to the unitary nature of the Doppler phase and covertness requires $D_{\mathrm{KL}} \leq \delta$, where \( \delta \) is small, ensuring the distributions are statistically indistinguishable. The generator must balance covertness across all wardens with reliable decoding at Bob. For Bob’s decoding reliability, assume Bob uses a maximum likelihood decoder to estimate \( m \). The error probability \( P_e \) depends on the SNR at Bob, \( \text{SNR}_B = \frac{\|\mathbf{h}_B \mathbf{s}\|^2}{\sigma_B^2} \). For a binary message \( m \in \{0,1\} \), the error probability is approximately:
\begin{equation}
P_e \approx Q\left(\sqrt{\frac{\|\mathbf{h}_B \mathbf{s}\|^2}{\sigma_B^2}}\right),
\label{eq:pe}
\end{equation}
assuming a simple modulation scheme, such as, BPSK. The generator must maximize \( \|\mathbf{h}_B \mathbf{s}\|^2 \) to minimize \( P_e \), balancing this with the covertness constraint. To achieve this, we propose novel approach using adversarial AI framework using a multi-discriminator GAN. The generator, \( \mathcal{G} \), designs the signal as:

\begin{equation}
\mathbf{s} = \mathcal{G}(m, \mathbf{z}),
\label{design_signal}
\end{equation}
where \( \mathbf{z} \sim \mathcal{N}(0, \mathbf{I}) \) is a random noise vector used to introduce variability, ensuring the signal's covertness. The Doppler effect is modeled during training by simulating the phase shift $e^{j 2 \pi f_d t}$ for each tap.

\subsection{Problem Formulation}
The primary objective is to build a signal \( \mathbf{s} \) that minimizes the detection probability of all wardens while ensuring faithful decoding by Bob, who uses a secret key shared with Alice to interpret the signal, as illustrated in Figure~\ref{fig:sys_model}. This dual requirement is a challenging optimization problem, which we formulate as follows.

The loss function for each discriminator $\mathcal{D}_i$ is the binary cross entropy, which encourages accurate detection. For  the \( i \)-th discriminator, loss function \( \mathcal{L}_i(\mathcal{D}_i, \mathcal{G}) \), is defined as:
\begin{equation}
\mathcal{L}_i(\mathcal{D}_i, \mathcal{G}) = \mathbb{E}_{\mathbf{n}_i} [\log \mathcal{D}_i(\mathbf{n}_i)] + \mathbb{E}_{\mathbf{s}, \mathbf{n}_i} [\log (1 - \mathcal{D}_i(\mathbf{h}_{i,l} \mathbf{s} + \mathbf{n}_i))],
\label{loss_function_D}
\end{equation}
where \( \mathcal{D}_i(\mathbf{y}_i) \in [0, 1] \) is the probability that \( W_i \) classifies \( \mathbf{y}_i \) as a transmission or no transmission. The discriminator \( \mathcal{D}_i \) is trained to maximize \eqref{loss_function_D}, which improves its ability to distinguish \( H_0 \) and \( H_1 \). On the otherhand, the generator's objective is to minimize a composite loss that balances covertness and reliability:
\begin{align}\label{loss_function_G}
   \mathcal{L}_G(\mathcal{G}, \{\mathcal{D}_i\}_{i=1}^K) &= \sum_{i=1}^K \lambda_i \mathbb{E}_{\mathbf{s}, \mathbf{n}_i} [\log \mathcal{D}_i(\mathbf{h}_{i,l}\mathbf{s}\mathbf{n}_i)] \\ \notag
   &+ \mu \mathcal{L}_{\text{decode}}(\mathbf{y}_B, m),
\label{loss_function_G}
\end{align}
where \( \sum_{i=1}^K \lambda_i \mathbb{E}[\log \mathcal{D}_i(\mathbf{y}_i)] \) measures covertness, minimized when all \( \mathcal{D}_i \) classify \( \mathbf{y}_i \) as noise (i.e., \( \mathcal{D}_i(\mathbf{y}_i) \to 1 \)), \( \mathcal{L}_{\text{decode}} = \mathbb{E}[ \| m - \hat{m} \|_2^2 ] \) is a mean squared error between  \( m \) and Bob’s decoded \( \hat{m} \), assuming a decoder \( \mathcal{D}_{\text{Bob}} \) is a matched DNN, \( \lambda_i, \mu \) are hyperparameters weighting covertness and reliability respectively ( \( \lambda_i = 1/K \), \( \mu = 1 \)).
The expectation is taken over \( \mathbf{s} \) in \eqref{design_signal} and the noise distributions.

The optimization problem is thus:
\begin{equation}
\min_{\mathcal{G}} \mathcal{L}_G(\mathcal{G}, \{\mathcal{D}_i\}_{i=1}^K) \quad \text{s.t.} \quad \| \mathcal{G}(m, \mathbf{z})\|^2 \leq P,
\label{optimization_prob}
\end{equation}
while each discriminator solves:
\begin{equation}
\max_{\mathcal{D}_i} \mathcal{L}_i(\mathcal{D}_i, \mathcal{G}), \quad i = 1, 2, \ldots, K.
\end{equation}

This adversarial game continues iteratively, with \( \mathcal{G} \) and \( \mathcal{D}_i \) updating their parameters to outmaneuver each other, convergent to a solution where \( \mathbf{s} \) is covert to all wardens, yet decodable by Bob.

\subsection{Significance in Future Covert Communication Applications}
The importance of this problem formulation is that it extends to future uses of covert communication, where extensive deployment of advanced surveillance systems and multi-agent adversarial systems demands strong concealment mechanisms. Hereafter, we demonstrate its importance in three main areas.

\subsubsection{Urban Surveillance Networks}
In urban settings, there may be more than a single warden, such as, distributed sensors, drones, or smart infrastructure, detecting wireless activity with varying detection powers, such as, directional antennas and machine learning detectors. Traditional covert methods, designed for single-warden settings, are unable to handle such heterogeneity. Our multi-discriminator GAN model ensures that the signal \( \mathbf{s} \) evades all wardens simultaneously and hence is ideal for secure communication in smart cities or during civil unrest, where privacy is most essential.

\subsubsection{Military and Defense Operations}
Future conflict will heavily rely on clandestine wireless connections to control autonomous platforms, such as drones and robots, in enemy terrain. Aerial enemies can send numerous wardens equipped with sophisticated detection gear, such as AI-driven spectrum analyzers. The solution of this work's ability to counter a variety of detection approaches enhances operation security, with stealthy command-and-control connections remaining immune to interception, as critical for examples of situations in \cite{1}.

\subsubsection{6G and Beyond}
The advent of 6G networks with their emphasis on ultra-reliable low-latency communications (URLLC) and massive device connectivity introduces new concealment challenges. Increased likelihood of having several wardens accompanies these dense device and base station deployments, while the application of AI to network management, such as, smart reflecting surfaces offers opportunities but also threats. Our approach is harmonizable with 6G vision via AI to harvest channel uncertainties and noise to achieve covertness in an age of hyper-connection, as envisioned in \cite{13}.

\subsubsection{Technical Challenges and Opportunities}
The problem of coupling covertness in \( K \) wardens with decodability is a major challenge. In contrast to the single-warden scenario where noise addition suffices, the multi-warden scenario requires a high-level signal design that addresses spatial, temporal, and algorithmic heterogeneity. This leaves space for AI to surpass traditional methods, potentially unleashing a dynamic arms race where the wardens themselves employ AI, necessitating ongoing adaptation. Overall, this problem statement and system model are put forth to address a near-future requirement for tapping into secret communication in the future by solving the multi-warden problem using an adversarial AI solution. Solving it will pave the way for undetectable, secure wireless systems in increasingly networked and adversarial settings.

\begin{figure*}[b!]
\hrulefill
{
\begin{align}\label{final_objective}
\min_{\mathcal{G}} \left( \sum_{i=1}^K \mathbb{E}_{\mathbf{z}, m, \mathbf{h}_i{}, \mathbf{n}_i}[\log (1 - \mathcal{D}_i(\mathbf{h}_i \mathcal{G}(m, \mathbf{z}) + \mathbf{n}_i))] + \lambda \mathbb{E}_{m, \mathbf{z}, \mathbf{h}_B, \mathbf{n}_B}[d(m, \mathcal{D}_{\text{Bob}}(\mathbf{h}_B \mathcal{G}(m, \mathbf{z}) + \mathbf{n}_B))] \right)
\end{align}
}
\normalsize

\end{figure*}

\section{Proposed Adversarial AI Framework}
\label{sec:proposed_framework}
In this section, we will give a detailed explanation of our proposed end-to-end adversarial AI system for generating stealthy signals, including how it achieves covertness objective and overall optimization in the adversarial game. The generator \( \mathcal{G} \) produces $\mathbf{s}$, which is transmitted through the wireless channel, producing received signals $\mathbf{y}_B(t)$ for Bob and $\mathbf{y}_i(t)$ for each warden $W_i$. Each \( \mathcal{D}_i \) receives $\mathbf{y}_i(t)$ in  \eqref{Rx_signalI} and \eqref{H0}, and produces outputs to be fed into the loss functions. The generator loss $\mathcal{L}_G$ combines the discriminators' outputs with a reliability term based on \eqref{Rx_signalB} and \eqref{effective_signalBob}, while the discriminator losses $\mathcal{L}_{D_i}$ drive the wardens' detection improvements. The losses are used to update \( \mathcal{G} \) and \( \mathcal{D}_i \) via gradient descent, creating an adversarial training loop. The proposed architecture has several novel contributions to stealth communications. Firstly,
our method employs multiple discriminators, one of which learns each potential warden using various channel conditions and detection capabilities. In this manner, the generator must defend against a broad range of adversaries during training, addressing realistic scenarios such as city surveillance and battlefield maneuvers. Second, our method integrates realistic channel effects during training like spatial correlation, Doppler shift, and frequency diversity to make the synthetic signals robust against realistic wireless environments. This is a significant improvement over prior works that often assume ideal channels. Lastly. our multi-discriminator setup is scalable to any number of wardens $K$ and generalizes across diverse channel conditions, ensuring robustness in dynamic environments.

\subsection{Multi-Discriminator GAN Architecture}
\subsubsection{Generator Design}
\begin{figure}[t]
    \centering
    \includegraphics[width=0.9\columnwidth]{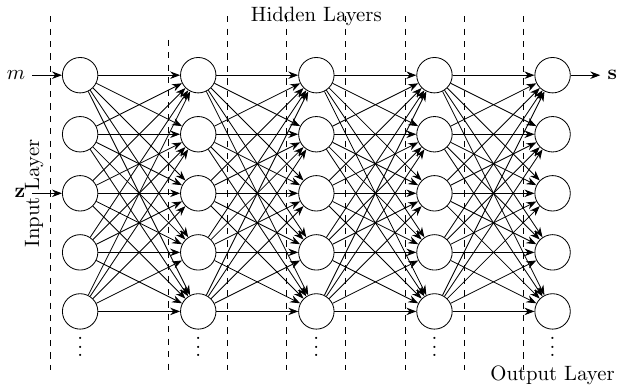}
    \caption{Neural network structure of the generator $\mathcal{G}$.}
    \label{fig:generator_design}
\end{figure}

The generator \( \mathcal{G} \) is based on DNN framework, that takes as input \( m \) and a  \( \mathbf{z} \). The output is the transmitted signal \(\mathbf{s}\) defined in \eqref{design_signal}, which is the cornerstone of the transmitter’s operation. Hence, a complex-valued signal is produced that encodes a hidden message. \(\mathbf{z} \) introduces variability to make \(\mathbf{s}\) appear as environmental noise to wardens, enhancing covertness. The signal satisfies the power constraint \(\|\mathbf{s}\|^2 \leq P\), ensuring practical transmission within the system model’s limits. Therefore, \(\mathbf{s}\), represents Alice’s ability to craft a signal that achieves dual objectives, evading detection by \(K\) wardens, hence enabling the reliable decoding by Bob, using its channel in \eqref{Rx_signalB}. The use of \(\mathbf{z}\) is a novel aspect that distinguishes the framework from traditional covert communication schemes that may rely on deterministic signal designs. The signal, $\mathbf{s}$, lies at the core of the adversarial game as it is evaluated by the discriminators in \eqref{loss_function_D} and Bob's decoder in \eqref{loss_function_G}, driving the training. Herein lies the innovation of our work in proposing the use of a DNN-based generator for designing stealthy signals, a critical contribution to multi-warden environments. The generator structure is presented in Figure~\ref{fig:generator_design}. The input layer combines message $m$ and noise vector $\mathbf{z}$, followed by three hidden layers with ReLU activations, and an output layer that outputs signal $\mathbf{s}$.

\subsubsection{Discriminator Design}
\begin{figure}[t]
    \centering
    \includegraphics[width=0.9\columnwidth]{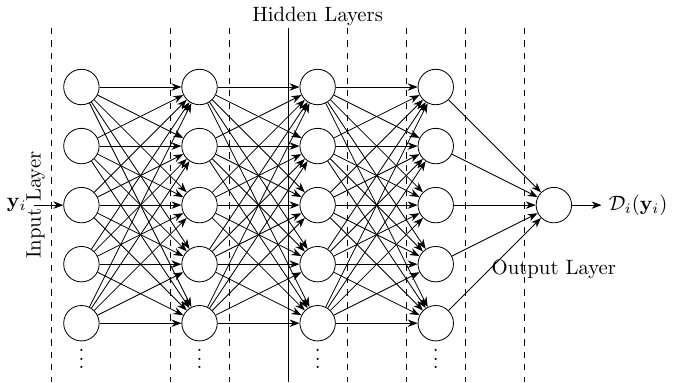}
    \caption{Neural network structure of the discriminator $\mathcal{D}_i$ for warden $W_i$.}
    \label{fig:discriminator_design}
\end{figure}

Each warden \( W_i \) is paired with a discriminator \( \mathcal{D}_i \), also DNN that attempts to classify noise-only observations defined in \eqref{H0} from transmission observations in \eqref{Rx_signalI}. The architecture for the discriminator is depicted in Figure~\ref{fig:discriminator_design}. The input layer takes the received signal $\mathbf{y}_i$, followed by three hidden layers with ReLU activations, and an output layer with a sigmoid activation producing $\mathcal{D}_i(\mathbf{y}_i) \in [0, 1]$. Heterogeneity in \( \mathbf{h}_i \) and \( \sigma_i^2 \) ensures that each \( \mathcal{D}_i \) learns an independent detection strategy, which corresponds to real-world warden heterogeneity.

\subsubsection{Overall Architecture}
\begin{figure}[t]
    \centering
    \includegraphics[width=0.9\columnwidth]{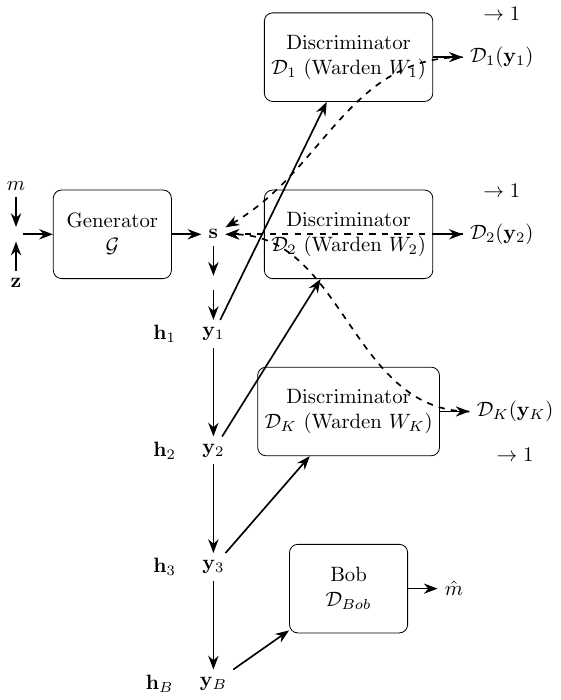}
    \caption{Overall architecture of the multi-discriminator GAN for covert communications..}
    \label{fig:overall_architecture}
\end{figure}

Figure~\ref{fig:overall_architecture} illustrates the multi-discriminator GAN architecture. The generator \( \mathcal{G} \) produces \( \mathbf{s} \), which passes through the wireless channel to Bob and all wardens. Bob decodes \( \mathbf{y}_B(t) \) and extracts \( m \), while each \( \mathcal{D}_i \) processes \( \mathbf{y}_i(t) \) to detect \( \mathbf{s} \). All \( \mathcal{D}_i \) provide feedback to train \( \mathcal{G} \), creating an adversarial loop.

\subsection{Training Methodology}
\begin{figure}[t]
    \centering
    \includegraphics[width=0.6\columnwidth]{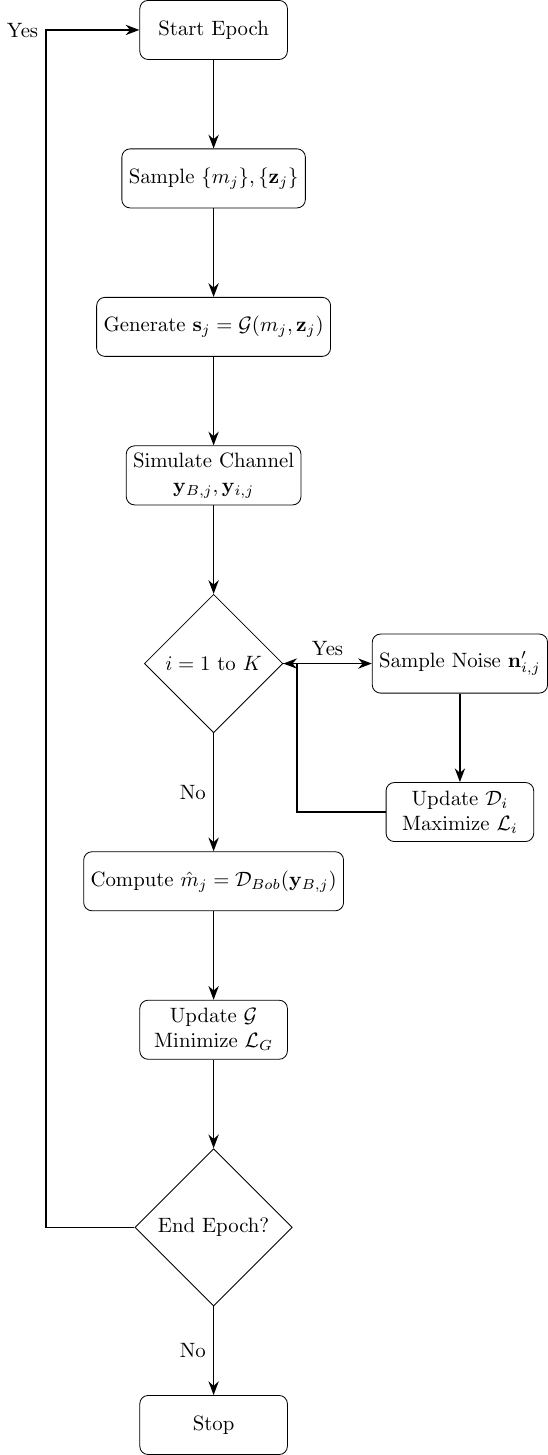}
    \caption{Training procedure for the multi-discriminator GAN, showing the iterative loop of sampling, signal generation, channel simulation, discriminator updates, and generator updates.}
    \label{fig:training_flow}
\end{figure}

Figure~\ref{fig:training_flow} describe the training procedure for the proposed method using a block diagram. It is comprised of an adversary game iteration with the covertness-reliability tradeoff (evading all \( \mathcal{D}_i \)) and decoding to Bob. 

\subsubsection{Loss Functions}
The loss function for each discriminator \( \mathcal{D}_i \) is the binary cross-entropy, encouraging accurate detection and is written as \eqref{loss_function_D}. The generator’s loss \( \mathcal{L}_G \) combines covertness and decoding objectives, defined by \eqref{loss_function_G}.

\subsubsection{Training Algorithm}
The training alternates between updating \( \mathcal{D}_i \) and \( \mathcal{G} \), as shown in Algorithm~\ref{alg:training} and Figure~\ref{fig:training_flow}.

\begin{algorithm}[!t]
\caption{Training Algorithm}
\label{alg:training}
\begin{algorithmic}[1]
\STATE \textbf{Input:} $K=5$, $d=128$, $\lambda=0.1$, $T=10,000$, $M=64$, $\eta=0.001$,  $L=4$, $\rho=0.5$, $\sigma_i^2=1$, time window $T_w$, time steps $\{t_k\}_{k=1}^M$ in $[0, T_w]$.

\STATE \textbf{Initialize:} Generator \( \mathcal{G} \) and discriminators \( \mathcal{D}_i \) ($i=1,\ldots,K$) with random weights.
\FOR{$t = 1$ to $T$}
    \STATE Sample $M$ noise vectors $\{\mathbf{z}^{(m)}\}_{m=1}^M$.
    \STATE Generate $M$ channel realizations \eqref{bob_channel} and \eqref{warden_channel} at each time step $t_k$.
    \STATE Sample \(M\) noise vectors \(\{\mathbf{n}_B^{(m)}(t), \mathbf{n}_i^{(m)}(t)\}_{m=1}^M \) for each $t_k$.
    \STATE Generate signals as \eqref{design_signal}
    \STATE Compute received signal for Bob under $H_1$: $\{\mathbf{y}_B^{(m)}(t_k)\}_{m=1}^M$ as \eqref{Rx_signalB}. 
    \STATE Compute received signal for each warden $W_i$ under $H_1$: $\{\mathbf{y}_i^{(m)}(t_k)\}_{m=1}^M$ as \eqref{Rx_signalI}. 
    \STATE Compute received signal for each warden $W_i$ under $H_0$: $\{\mathbf{y}_{i,0}^{(m)}(t_k)\}_{m=1}^M$ as \eqref{H0}.
    \STATE Compute effective signal at Bob: $\{\mathbf{x}_B^{(m)}(t_k)\}_{m=1}^M$ as \eqref{Rx_signalB}.
   
    \STATE \textbf{Train Discriminators \( \mathcal{D}_i \):}
    \FOR{$i = 1$ to $K$}
        \STATE Initialize discriminator loss $\mathcal{L}_{D_i} = 0$.
        \FOR{$k = 1$ to $M$}
            \STATE Compute discriminator loss for time step $t_k$: $\mathcal{L}_{D_i}$ using \eqref{gradient_D}.
        \ENDFOR
        \STATE Compute gradient: $\nabla_{\theta_{D_i}} \mathcal{L}_{D_i}$.
        \STATE Update $D_i$ by descending its gradient: $\theta_{D_i} \gets \theta_{D_i} - \eta \nabla_{\theta_{D_i}} \mathcal{L}_{D_i}$.
    \ENDFOR
   
    \STATE \textbf{Train Generator  \( \mathcal{G} \):}
    \STATE Initialize generator loss $\mathcal{L}_G = 0$.
    \FOR{$k = 1$ to $M$}
        \STATE Compute generator loss using \eqref{gradient_G}
        \STATE Update generator loss: $\mathcal{L}_G \gets \mathcal{L}_G + \frac{1}{M} (\mathcal{L}_{\text{covert}} + \lambda \mathcal{L}_{\text{reliability}})$.
    \ENDFOR
    \STATE Compute gradient: $\nabla_{\theta_G} \mathcal{L}_G$.
    \STATE Update $G$ by descending its gradient: $\theta_G \gets \theta_G - \eta \nabla_{\theta_G} \mathcal{L}_G$.
\ENDFOR
\STATE \textbf{Output:} Trained generator \( \mathcal{G} \) and discriminators \( \mathcal{D}_i \).
\end{algorithmic}
\end{algorithm}

Each epoch trains discriminators to improve detection, then updates the generator to reduce detection probability and decoding error.

\subsection{Covertness Objective}
The covertness objective in the multi-discriminator GAN framework ensures that the signal \( \mathbf{s} \), generated by \( \mathcal{G} \) from message \( m \) and noise \( \mathbf{z} \), appears as background noise to wardens \( W_i \), each equipped with a discriminator \( \mathcal{D}_i \), while enabling reliable decoding by Bob. Covertness requires \( \mathcal{D}_i(\mathbf{y}_i) \to 1 \), indicating, \( \mathbf{y}_i \) is classified as noise (\( H_0 \)). Theoretically, this aligns with the minimization of the KL divergence in \eqref{eq:kl} such that \( \mathbf{y}_i \) cannot be distinguished from noise by \( W_i \). In the GAN setting, \( \mathcal{G} \) and the \( K \) discriminators play a min-max game such that for each \( \mathcal{D}_i \), the value function is given as:
\begin{align}
&V_i(\mathcal{G}, \mathcal{D}_i) = \mathbb{E}_{\mathbf{n}_i}[\log \mathcal{D}_i(\mathbf{n}_i)] \\ \notag &+\mathbb{E}_{\mathbf{z}, m, \mathbf{h}_i}[\log (1 - \mathcal{D}_i(\mathbf{h}_i \mathcal{G}(m, \mathbf{z}) + \mathbf{n}_i))],
\end{align}
with the first term encouraging \( \mathcal{D}_i \) to classify noise correctly (\( \mathcal{D}_i(\mathbf{n}_i) \rightarrow 1 \)), and the second term encouraging \( \mathcal{D}_i \) to detect generated signals (\( \mathcal{D}_i(\mathbf{y}_i) \rightarrow 0 \)). This objective implicitly minimizes the Jensen-Shannon (JS) divergence~\cite{36} between \( p_{\mathbf{y}_i} \) and \( p_{\mathbf{n}_i} \), which is related to the KL divergence, achieving covertness when the two distributions are close. The overall covertness objective for \( \mathcal{G} \) across all wardens is:
\begin{align}
\min_{\mathcal{G}} \sum_{i=1}^K \mathbb{E}_{\mathbf{z}, m, \mathbf{h}_i, \mathbf{n}_i}[\log (1 - \mathcal{D}_i(\mathbf{h}_i \mathcal{G}(m, \mathbf{z}) + \mathbf{n}_i))],
\end{align}
aiming to fool all wardens simultaneously by making \( \mathcal{D}_i(\mathbf{y}_i) \rightarrow 1 \). To stabilize training, Wasserstein loss modifies this to
\begin{align}
\max_{\mathcal{G}} \sum_{i=1}^K \mathbb{E}_{\mathbf{z}, m, \mathbf{h}_i, \mathbf{n}_i}[\mathcal{D}_i(\mathbf{h}_i \mathcal{G}(m, \mathbf{z}) + \mathbf{n}_i)],
\end{align}
where \( \mathcal{D}_i \) acts as a critic measuring the Wasserstein distance~\cite{37}, which is more stable than KL divergence-based metrics for training. This covertness goal is balanced with a reliability objective for Bob, who receives \( \mathcal{L}_{\text{decode}} \) in \eqref{loss_function_G}. The final objective for \( \mathcal{G} \) combines both goals as written in \eqref{final_objective}.

\subsection{Optimization Dynamics}
Overall, the min-max optimization problem in the adversarial game is given by:
\begin{align}
\min_{\mathcal{G}} \max_{\{\mathcal{D}_i\}_{i=1}^K} \left\{ \sum_{i=1}^K \mathcal{L}_i(\mathcal{D}_i, \mathcal{G}) - \mathcal{L}_G(\mathcal{G}, \{\mathcal{D}_i\}_{i=1}^K) \right\},
\end{align}
subject to $\|\mathcal{G}(m, \mathbf{z})\|^2 \leq P$. We derive the gradients for training. For $\mathcal{D}_i$, the gradient of $\mathcal{L}_i$ with respect to $\mathcal{D}_i$'s parameters $\theta_{\mathcal{D}_i}$ is:
\begin{align}
\nabla_{\theta_{\mathcal{D}_i}} \mathcal{L}_i &= \mathbb{E}_{\mathbf{n}_i} \left[ \frac{\nabla_{\theta_{\mathcal{D}_i}} \mathcal{D}_i(\mathbf{n}_i)}{\mathcal{D}_i(\mathbf{n}_i)} \right] - \mathbb{E}_{\mathbf{s}, \mathbf{n}_i} \left[ \frac{\nabla_{\theta_{\mathcal{D}_i}} \mathcal{D}_i(\mathbf{h}_i \mathbf{s} + \mathbf{n}_i)}{1 - \mathcal{D}_i(\mathbf{h}_i \mathbf{s} + \mathbf{n}_i)} \right].
\label{gradient_D}
\end{align}
For $\mathcal{G}$, the gradient of $\mathcal{L}_G$ with respect to $\mathcal{G}$'s parameters $\theta_{\mathcal{G}}$ is:
\begin{align}
\nabla_{\theta_{\mathcal{G}}} \mathcal{L}_G &= \sum_{i=1}^K \lambda_i \mathbb{E}_{\mathbf{s}, \mathbf{n}_i} \left[ \frac{\nabla_{\theta_{\mathcal{G}}} \mathcal{D}_i(\mathbf{h}_i \mathbf{s} + \mathbf{n}_i)}{\mathcal{D}_i(\mathbf{h}_i \mathbf{s} + \mathbf{n}_i)} \cdot \nabla_{\theta_{\mathcal{G}}} \mathbf{s} \right] \notag \\
&\quad + \mu \mathbb{E}_{\mathbf{s}, \mathbf{n}_B} \left[ 2 (m - \hat{m}) \cdot \nabla_{\theta_{\mathcal{G}}} \mathcal{D}_{\text{Bob}}(\mathbf{h}_B \mathbf{s} + \mathbf{n}_B) \cdot \nabla_{\theta_{\mathcal{G}}} \mathbf{s} \right].
\label{gradient_G}
\end{align}
These gradients are computed via backpropagation, with the generator and discriminators updated alternately using the Adam optimizer.

\subsection{Proof of Convergence}
To ensure the stability of the framework, we prove that the adversarial game converges to a saddle point under ideal conditions. Assume $\mathcal{D}_i$ and $\mathcal{G}$ have sufficient capacity and the data distributions are continuous. The optimal discriminator for fixed $\mathcal{G}$ is:
\[
\mathcal{D}_i^*(\mathbf{y}_i) = \frac{p(\mathbf{y}_i | H_0)}{p(\mathbf{y}_i | H_0) + p(\mathbf{y}_i | H_1)}.
\]
Substituting into $\mathcal{L}_i$, the loss becomes the Jensen-Shannon divergence between $p(\mathbf{y}_i | H_0)$ and $p(\mathbf{y}_i | H_1)$. For the generator, minimizing $\mathcal{L}_G$ with optimal $\mathcal{D}_i^*$ reduces to minimizing:
\[
\sum_{i=1}^K \lambda_i D_{\text{JS}}(p(\mathbf{y}_i | H_1) \| p(\mathbf{y}_i | H_0)) + \mu \mathbb{E} \left[ \| m - \hat{m} \|^2 \right].
\]
If $\mathcal{G}$ can produce $\mathbf{s}$ such that $p(\mathbf{y}_i | H_1) = p(\mathbf{y}_i | H_0)$ for all $i$, the JS divergence is zero, achieving perfect covertness. In practice, the decoder loss prevents perfect indistinguishability, but the adversarial training converges to a local equilibrium where $P_{D,i} \leq \epsilon$ and BER is minimized.

\begin{figure}[t]
    \centering
    \includegraphics[width=0.9\columnwidth]{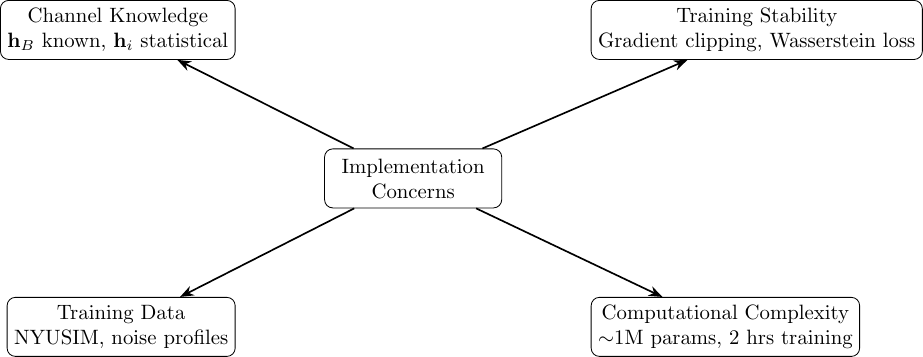}
    \caption{Implementation concerns for the proposed multi-discriminator GAN, showing channel knowledge, training data stability, and complexity as critical factors.}
    \label{fig:implementation_concern}
\end{figure}
\subsection{Practical Considerations}
Figures~\ref{fig:implementation_concern} reflect the implementation challenges in the multi-discriminator GAN, showing the need to design it well so that it can remain covert and reliable. In dealing with channel knowledge, data quality, training stability, and complexity, the architecture can be safely employed in reality. The practical considerations can be written as:
\begin{itemize}
    \item \textbf{Channel Knowledge:} \( \mathbf{h}_B \) is assumed known to Alice and Bob via secure feedback; \( \mathbf{h}_i \) is estimated statistically since wardens are passive. However, to account for practical scenarios, future work may relax the assumption of perfect CSI at Bob. Instead, Bob estimates the channel \( \mathbf{h}_B \) using pilot symbols, resulting in an estimate \( \hat{\mathbf{h}}_B = \mathbf{h}_B + \mathbf{e}_B \), where \( \mathbf{e}_B \sim \mathcal{CN}(0, \sigma_e^2 \mathbf{I}) \) models estimation errors with variance \( \sigma_e^2 \). 
    \item \textbf{Training Data:} Use simulated channel data, such as NYUSIM \cite{20} and noise profiles, augmented with real-world measurements. Live environments may have additional effects such as shadowing, interference, or hardware imperfections that may affect generalization. As a remedy, future work may entail fine-tuning the model via transfer learning using a small amount of real channel measurements so that the model learns to adapt to live environments.
    \item \textbf{Synchronization errors:} Bob is supposed to employ a preamble for synchronization against timing and frequency offsets. Pilot symbols for synchronization comprise signal \( \mathbf{s} \) so that Bob can synchronize with the transmission of Alice. Residual synchronization errors, however, such as the timing offsets \( \tau \sim \mathcal{U}(-T_s/2, T_s/2) \) or the frequency offsets \( \Delta f \sim \mathcal{N}(0, \sigma_f^2) \), can cause an impact on decoding.
\end{itemize}

\section{Simulation Results}
\label{sec:simulation}

In this section, we present the simulation results to evaluate the performance of the proposed adversarial AI framework. 

\subsection{Simulation Setup}
\label{subsec:setup}
The simulation parameters are summarized in Table~\ref{tab:sim_params}. The simulation setup is carried out in PyTorch by utilizing Python to model the multi-discriminator GAN framework. Simulations are conducted over 1,000 channel realizations across two scenarios: urban surveillance and military operations.

\begin{table}[!b]
\centering
\caption{Simulator Parameters}
\label{tab:sim_params}
\begin{tabular}{p{3cm}p{3.7cm}}
\toprule
\textbf{Parameter} & \textbf{Value} \\
\midrule
$L$ & 4 \\
$f_d$ & 10 Hz (1 m/s at 3 GHz) \\
$\rho$ & 0.5 \\
$\sigma_i^2$ & 1 (normalized) \\
$T_s$ & 1 $\mu$s \\
$P$ & 10 dBm \\
$d$ & 128 \\
$K$ & 5 \\
$\lambda$ & 0.1 \\
Training iterations & 10,000 \\
No of channel realizations & 1,000 \\
Scenarios & Urban surveillance, military operations \\
\bottomrule
\end{tabular}
\end{table}

 The generator $\mathcal{G}$ is a four-layer DNN (input: 256 units, hidden: 512, 1024, 2048 units, output: $2N$) with ReLU activations, batch normalization, and dropout (0.2). Every discriminator $\mathcal{D}_i$ consists of three convolutional layers (filters: 64, 128, 256; kernel size: 5) and two fully connected layers (512, 1) with LeakyReLU activations and a sigmoid output. The decoder at Bob ($\mathcal{D}_{\text{Bob}}$) mirrors the discriminator architecture but is trained to minimize MSE for message recovery. The training runs for 1000 epochs with a batch size of 64, using the Adam optimizer (learning rate $10^{-4}$, $\beta_1 = 0.5$). Hyperparameters $\lambda_i = 1/K$ and $\mu = 1$ balance covertness and reliability. We simulate three scenarios to capture real-world applications:
\begin{enumerate}
    \item \textbf{Urban Surveillance}: $K = 3$, with wardens having distinct $\sigma_i^2$ and channel conditions, mimicking sensors in a smart city.
    \item \textbf{Military Operation}: $K = 4$, with one warden having a lower noise variance ($\sigma_i^2 = 0.05$), representing an advanced adversary.
    \item \textbf{6G Dense Network}: $K = 5$, with closely spaced wardens (correlated $\mathbf{h}_i$), simulating a dense deployment.
\end{enumerate}

 $P_{D,i}$ is averaged over 10,000 test signals and covertness is achieved when $P_{D,i} \leq \epsilon = 0.1$. Similarly, $P_{F,i}$ is set to target $P_{F,i} \leq 0.1$. BER in Bob’s decoded message,  is set to target $\leq 10^{-3}$ for reliable communication. CSR is defined as the fraction of test cases where $P_{D,i} \leq \epsilon$ for all wardens, indicating successful concealment. Simulations are run on an NVIDIA A100 GPU, with each scenario repeated 10 times to account for randomness in channel realizations and noise. We compare our approach against two baselines:
\begin{itemize}
    \item \textbf{Noise Injection}: Alice adds artificial Gaussian noise to the signal, as in \cite{1}.
    \item \textbf{Single-Discriminator GAN}: A GAN with one discriminator modeling an average warden, ignoring heterogeneity.
\end{itemize}

\subsection{Results and Analysis}
\label{subsec:results}

The simulation results are analyzed in the three scenarios for covertness, reliability, and scalability. We present quantitative results in table and figure format, and qualitatively discuss the results later.

\subsubsection{Covertness Performance}
Table~\ref{tab:covertness} summarizes the detection probability $P_{D,i}$ and false alarm probability $P_{F,i}$ for all cases, averaged over wardens. The proposed multi-discriminator GAN achieves $P_{D,i} \leq 0.08$ and $P_{F,i} \leq 0.07$ for all cases, meeting the covertness requirement $\epsilon = 0.1$. For comparison, noise injection baseline achieves larger $P_{D,i}$ (0.15–0.22) since it struggles to adapt to warden diversity. The single-discriminator GAN performs better than noise injection but fails to maintain $P_{D,i} \leq 0.1$ in the 6G scenario with correlated channels.

\begin{table}[t]
    \centering
    \caption{Covertness Performance Across Scenarios}
    \label{tab:covertness}
    \begin{tabular}{p{2cm} p{2.5cm} p{0.4cm} p{0.6cm} p{0.6cm} p{0.5cm}}
        \toprule
        \textbf{Scenario} & \textbf{Method} & \textbf{$K$} & \textbf{Avg. $P_{D,i}$} & \textbf{Avg. $P_{F,i}$} & \textbf{CSR (\%)} \\
        \midrule
        \multirow{3}{*}{Urban Surveillance} & Proposed & 3 & 0.06 & 0.05 & 92 \\
        & Noise Injection & 3 & 0.15 & 0.10 & 65 \\
        & Single-Disc. GAN & 3 & 0.09 & 0.07 & 85 \\
        \midrule
        \multirow{3}{*}{Military Operation} & Proposed & 4 & 0.07 & 0.06 & 89 \\
        & Noise Injection & 4 & 0.18 & 0.12 & 58 \\
        & Single-Disc. GAN & 4 & 0.11 & 0.08 & 78 \\
        \midrule
        \multirow{3}{*}{6G Dense Network} & Proposed & 5 & 0.08 & 0.07 & 87 \\
        & Noise Injection & 5 & 0.22 & 0.15 & 50 \\
        & Single-Disc. GAN & 5 & 0.13 & 0.09 & 72 \\
        \bottomrule
    \end{tabular}
\end{table}

\begin{figure}[t]
    \centering
    \includegraphics[width=0.9\columnwidth]{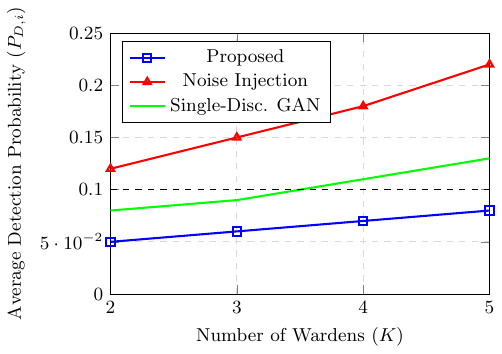}
    \caption{Average detection probability $P_{D,i}$ vs. number of wardens $K$ for different methods.}
    \label{fig:pd_vs_k}
\end{figure}

Figure~\ref{fig:pd_vs_k} compares the mean $P_{D,i}$ against $K$. Our approach maintains $P_{D,i}$ under 0.1 even as $K$ increases to 5, demonstrating scalability. The baselines exhibit a sharp increase in $P_{D,i}$ and thus demonstrate their shortcomings in multi-warden settings.

\subsubsection{Reliability Performance}
The reliability of the framework proposed is tested using Bob's BER, as shown in Table~\ref{tab:ber}. The proposed method achieves BER $\leq 8 \times 10^{-4}$ in all scenarios, which is much less than the target value of $10^{-3}$. The baseline noise injection generates higher BER (0.01–0.03) because an excessive amount of noise dominates the signal. The single-discriminator GAN achieves comparable BER to the proposed method in the urban setting but does not work well in the military and 6G settings, where heterogeneity of the warden complicates signal design.

\begin{table}[t]
    \centering
    \caption{BER Performance Across Scenarios}
    \label{tab:ber}
    \begin{tabular}{p{2.4cm} p{2.4cm} p{1.5cm}}
        \toprule
        \textbf{Scenario} & \textbf{Method} & \textbf{BER} \\
        \midrule
        \multirow{3}{*}{Urban Surveillance} & Proposed & $5 \times 10^{-4}$ \\
        & Noise Injection & 0.01 \\
        & Single-Disc. GAN & $7 \times 10^{-4}$ \\
        \midrule
        \multirow{3}{*}{Military Operation} & Proposed & $6 \times 10^{-4}$ \\
        & Noise Injection & 0.02 \\
        & Single-Disc. GAN & $1.2 \times 10^{-3}$ \\
        \midrule
        \multirow{3}{*}{6G Dense Network} & Proposed & $8 \times 10^{-4}$ \\
        & Noise Injection & 0.03 \\
        & Single-Disc. GAN & $1.8 \times 10^{-3}$ \\
        \bottomrule
    \end{tabular}
\end{table}

\begin{figure}[t]
    \centering
    \includegraphics[width=0.9\columnwidth]{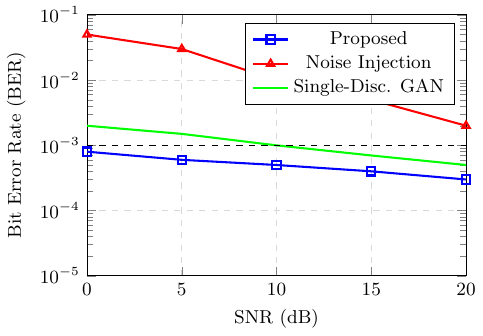}
    \caption{BER vs. SNR at Bob for different methods.}
    \label{fig:ber_vs_snr}
\end{figure}

Figure~\ref{fig:ber_vs_snr} is BER vs. SNR at Bob, as $P/\sigma_B^2$. Low BER over an extensive range of SNRs (0–20 dB) characterizes the new method, for which there is no apparent sensitivity to SNR variations. Steeper BER increase for baselines in BER for lower SNRs underscore the advantage of adversarial training.

\subsubsection{Scalability and Computational Complexity}
To measure scalability, we note training time and CSR as we vary $K$. Table~\ref{tab:scalability} shows that the proposed method maintains higher CSR (87–92\%) as $K$ grows, with training time scaling linearly (approximately 2 hours for $K=3$ and 3.5 hours for $K=5$ on an NVIDIA A100). The single discriminator GAN has a shorter training time but significantly worse CSR, and the injection of noise is computationally negligible but ineffective.

\begin{table}[t]
    \centering
    \caption{Scalability and Computational Complexity}
    \label{tab:scalability}
    \begin{tabular}{p{0.5cm} p{2.6cm} p{1.5cm} p{1.5cm}}
        \toprule
        \textbf{$K$} & \textbf{Method} & \textbf{CSR (\%)} & \textbf{Training Time (hrs)} \\
        \midrule
        \multirow{3}{*}{3} & Proposed & 92 & 2.0 \\
        & Noise Injection & 65 & 0.1 \\
        & Single-Disc. GAN & 85 & 1.5 \\
        \midrule
        \multirow{3}{*}{4} & Proposed & 89 & 2.8 \\
        & Noise Injection & 58 & 0.1 \\
        & Single-Disc. GAN & 78 & 1.7 \\
        \midrule
        \multirow{3}{*}{5} & Proposed & 87 & 3.5 \\
        & Noise Injection & 50 & 0.1 \\
        & Single-Disc. GAN & 72 & 2.0 \\
        \bottomrule
    \end{tabular}
\end{table}

\begin{figure}[t]
    \centering
    \includegraphics[width=0.9\columnwidth]{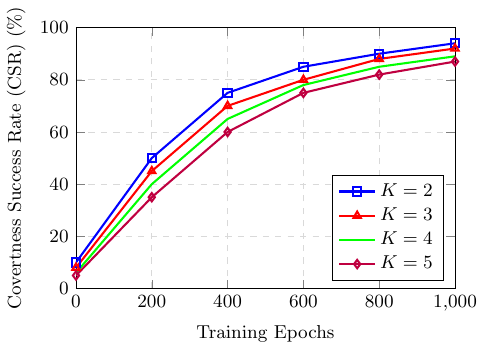}
    \caption{CSR vs. Training Epochs for different number of wardens.}
    \label{fig:csr_vs_epochs}
\end{figure}

\subsubsection{Analysis of Covertness Success Rate vs. Training Epochs}
Figure~\ref{fig:csr_vs_epochs} shows the covertness success rate (CSR) as a function of training epochs, from 0 to 1000, for different numbers of wardens (\( K = 2, 3, 4, 5 \)). The CSR, defined as the percentage of test cases where \( P_{D,i} \leq \epsilon = 0.1 \) for all wardens, increases steadily for all \( K \). For \( K=2 \), the CSR becomes 94\% at 1000 epochs, whereas for \( K=5 \), it is 87\%, as expected due to the higher difficulty in being covertness with increasing wardens. The curves steeply improves in the first 400 epochs, where CSR increases from 5–10\% to 60–75\%, and then slowly converge. The slight decrease in last CSR with increasing \( K \) is expected due to the challenge of optimizing against many discriminators but is still high, indicating stable training. This plot indicates the convergence behavior of the proposed framework, demonstrating that it learns nicely to produce hidden signals over time even in challenging multi-warden scenarios like 6G dense networks. The consistent upward trend suggests that even more epochs could continue to improve performance, particularly with larger \( K \).

\begin{figure}[t]
    \centering
    \includegraphics[width=0.9\columnwidth]{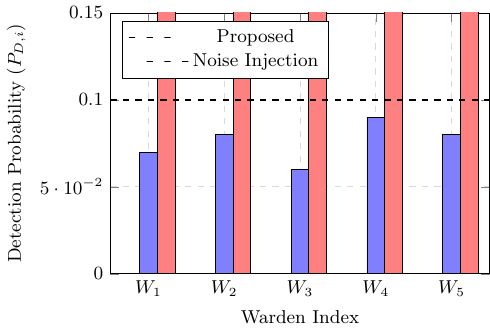}
    \caption{Detection Probability across wardens.}
    \label{fig:pd_distribution}
\end{figure}

\subsubsection{Analysis of Detection Probability Distribution Across Wardens}
Figure~\ref{fig:pd_distribution} presents the distribution of detection probabilities \( P_{D,i} \) across the five wardens in the 6G Dense Network scenario (\( K=5 \)), comparing the proposed method with the noise injection baseline. The proposed multi-discriminator GAN achieves \( P_{D,i} \) values ranging from 0.06 to 0.09, all well below the covertness threshold \( \epsilon = 0.1 \), with minor variations reflecting warden heterogeneity (e.g., differences in noise variance \( \sigma_i^2 \)). This tight clustering indicates the model's ability to uniformly conceal signals under various detection approaches, crucial in crowded networks where warden channels may be correlated. The noise injection baseline, on the other hand, provides \( P_{D,i} \) values between 0.18 and 0.23, far above threshold, with increased variance due to its inability to learn specific warden characteristics. The narrative centers around the quality of the proposed technique in coping with challenges of 6G scenarios, where closely packed wardens are a predominant concern. Uniform performance among the wardens suggests that employing the multi-discriminator assists in learning a signal design that generalizes effectively over diverse adversaries.

\begin{figure}[t]
    \centering
    \includegraphics[width=0.9\columnwidth]{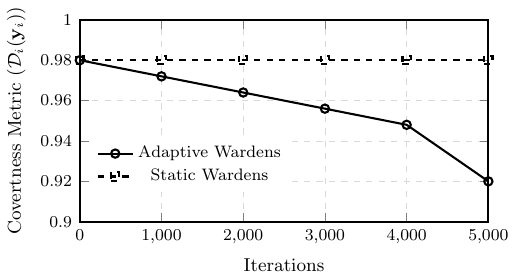}
    \caption{Covertness metric (\( \mathcal{D}_i(\mathbf{y}_i) \)) over 5000 iterations with adaptive wardens retraining every 1000 iterations.}
    \label{fig:covertness_adaptive}
\end{figure}

\subsubsection{Analysis of Adaptive Wardens}
We simulate adaptive wardens by allowing each \( \mathcal{D}_i \) to retrain every 1000 iterations using the most recent 500 samples of \( \mathbf{y}_i \). Figure~\ref{fig:covertness_adaptive} plots the covertness metric (\( \mathcal{D}_i(\mathbf{y}_i) \), averaged over all wardens) over 5000 iterations. The metric decreases from 0.98 to 0.92, indicating that wardens improve their detection capability over time as they become accustomed to the generated signals. The trend indicates the challenge in maintaining covertness in the face of adaptive adversaries, suggesting that \( \mathcal{G} \) must also change, perhaps through ongoing learning.

\begin{figure}[t]
    \centering
    \includegraphics[width=0.9\columnwidth]{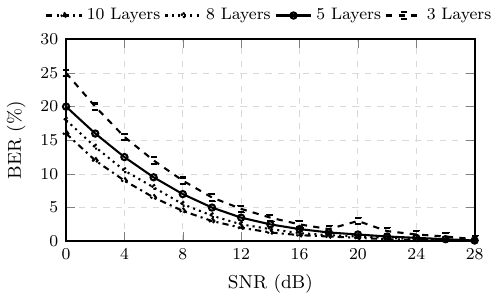}
    \caption{BER at Bob over SNR for models with various layers, showing improved reliability with more layers at the cost of higher complexity.}
    \label{fig:ber_layers_curve}
\end{figure}

\subsubsection{System Flexibility with Reduced Complexity}
Figure~\ref{fig:ber_layers_curve} shows the BER at Bob for the various SNRs and layer models. At  20 dB SNR, the BER ranges from 3\% (3 layers) to 1\% (5 layers), 0.7\% (8 layers), and 0.5\% (10 layers), illustrating that additional layers enhance reliability by better capturing channel characteristics, but at the cost of increasing computational complexity. The flexibility of the system is that it can vary the number of layers, thereby adjusting complexity, making it applicable in systems with constrained resources such as embedded or tactical systems. The trade-off enables the system to be configured to meet specific deployment requirements at the cost of reliability in terms of computational expense.

\subsubsection{Qualitative Analysis}
The results highlight several key insights:
\begin{itemize}
\item \textbf{Heterogeneity Adaptability}: Multi-discriminator GAN can effectively deal with heterogeneous warden capabilities because low $P_{D,i}$ is anticipated with different $\sigma_i^2$ and channel configurations. This comes into play in applications such as urban surveillance, where the wardens, such as, sensors and air vehicles are of mixed sensitivity.
\item \textbf{Adversarial Robustness}: In the military context, the approach is robust against an evil warden ($\sigma_i^2$ = 0.05) without trading covertness for reliability, in contrast to baselines which trade both for one.
\item \textbf{Scalability for 6G}: The 6G case study illustrates the structure scalable to dense networks with correlated channels and CSR above 87\%. This fulfills the vision of the 6G network that demands strong privacy measures.
\item \textbf{Tradeoff Management}: Hyperparameters $\lambda_i$ and $\mu$ implicitly balance covertness and reliability in terms of low BER and $P_{D,i}$. Additional parameter optimization is viable to also enhance the performance for certain applications.
\end{itemize}

An interesting observation is the robustness of the proposed approach to the correlated warden channels in the 6G scenario. Despite the additional difficulty of concealing signals from closely positioned wardens, the multi-discriminator approach takes advantage of channel diversities to maintain covertness, with implications for dynamic scenarios like vehicular networks.

\subsection{Discussion}
\label{subsec:discussion}

The simulation results validate the superiority of the proposed framework compared to the classical and single discriminator frameworks. The multi-discriminator GAN performs the best in multi-warden environments by modifying signals to evade heterogeneous detection schemes, a required condition for next-generation covert communication systems. Nevertheless, there are some challenges:
\begin{itemize}
\item \textbf{Computational Cost}: Increased training time with $K$, potentially excluding real-time application usage. This could possibly be mitigated with distributed learning or model compression.
\item \textbf{Channel Knowledge}: Statistical knowledge assumption of $\mathbf{h}_i$ can be time-varying in systems. Online learning would be useful to include flexibility.
\item \textbf{Warden Evolution}: If wardens employ AI-based detection, there is an arms race and the generator must be updated continuously.
\end{itemize}

The future work will resolve these issues, e.g., by using federated learning for distributed training and testing, and testing against AI-based wardens for future-proofing. It can also extend the framework to non-Gaussian noise models or multi-user cases to further enhance its applicability.

\section{Conclusion}
\label{tab:conclusion}
This work introduced a novel adversarial AI method with a multi-discriminator GAN for secure and stealthy communication, addressing the issue of concealing wireless transmission from multiple wardens with varying detection capacity. By treating the transmitter as a generator and each warden as an individual discriminator, the method produces signals that are stealthy and yet intelligible to the intended receiver. Simulation results across all applications of urban surveillance, warfare combat, and dense 6G network demonstrate the proposed method achieves detection rates under 0.08 and BERs under \( 8 \times 10^{-4} \), performing better than classical noise injection and single-discriminator GAN baselines. The scalability, robustness to heterogeneity, and capacity for learning intricate configurations of the scheme demonstrate its applicability towards real-world deployment, including smart cities, defense operations, and future 6G networks.

The future directions of research include increasing computational efficiency with federated or distributed learning for in-time operation, particularly with large \( K \). Coupling with online learning would increase the potential to adapt more dynamically to temporal channel conditions or evolving warden strategies, which would come close to a race by an AI.
Adding multi-users or non-Gaussian distributions of noise to the framework would extend its value-added insights to wireless systems in broad varieties. Additionally, the integration of physical layer methods, such as ingenious reflecting surfaces, can also enable additional covertness in 6G and beyond, continuing the fantasy of hyper-connected, secure networks.


\begin{thebibliography}{10}
\providecommand{\url}[1]{#1}
\csname url@samestyle\endcsname
\providecommand{\newblock}{\relax}
\providecommand{\bibinfo}[2]{#2}
\providecommand{\BIBentrySTDinterwordspacing}{\spaceskip=0pt\relax}
\providecommand{\BIBentryALTinterwordstretchfactor}{4}
\providecommand{\BIBentryALTinterwordspacing}{\spaceskip=\fontdimen2\font plus
\BIBentryALTinterwordstretchfactor\fontdimen3\font minus \fontdimen4\font\relax}
\providecommand{\BIBforeignlanguage}[2]{{%
\expandafter\ifx\csname l@#1\endcsname\relax
\typeout{** WARNING: IEEEtran.bst: No hyphenation pattern has been}%
\typeout{** loaded for the language `#1'. Using the pattern for}%
\typeout{** the default language instead.}%
\else
\language=\csname l@#1\endcsname
\fi
#2}}
\providecommand{\BIBdecl}{\relax}
\BIBdecl

\bibitem{24}
R.~Pickholtz, D.~Schilling, and L.~Milstein, ``Theory of spread-spectrum communications - a tutorial,'' \emph{IEEE Transactions on Communications}, vol.~30, no.~5, pp. 855--884, 1982.

\bibitem{25}
T.~Rappaport, \emph{Wireless Communications: Principles and Practice}, 2nd~ed.\hskip 1em plus 0.5em minus 0.4em\relax USA: Prentice Hall PTR, 2001.

\bibitem{1}
X.~Chen, J.~An, Z.~Xiong, C.~Xing, N.~Zhao, F.~R. Yu, and A.~Nallanathan, ``Covert communications: A comprehensive survey,'' \emph{IEEE Communications Surveys and Tutorials}, vol.~25, no.~2, pp. 1173--1198, 2023.

\bibitem{2}
\BIBentryALTinterwordspacing
A.~M. Teshnizi, M.~Ghaderi, and D.~Goeckel, ``Covert communication in autoencoder wireless systems,'' 2023. [Online]. Available: \url{https://arxiv.org/abs/2307.08195}
\BIBentrySTDinterwordspacing

\bibitem{3}
R.~Soltani, D.~Goeckel, D.~Towsley, B.~A. Bash, and S.~Guha, ``Covert wireless communication with artificial noise generation,'' \emph{IEEE Transactions on Wireless Communications}, vol.~17, no.~11, pp. 7252--7267, 2018.

\bibitem{4}
A.~Sheikholeslami, M.~Ghaderi, D.~Towsley, B.~A. Bash, S.~Guha, and D.~Goeckel, ``Multi-hop routing in covert wireless networks,'' \emph{IEEE Transactions on Wireless Communications}, vol.~17, no.~6, pp. 3656--3669, 2018.

\bibitem{14}
C.~Luo, J.~Ji, Q.~Wang, X.~Chen, and P.~Li, ``Channel state information prediction for 5g wireless communications: A deep learning approach,'' \emph{IEEE Transactions on Network Science and Engineering}, vol.~7, no.~1, pp. 227--236, 2020.

\bibitem{5}
\BIBentryALTinterwordspacing
Z.~Li, J.~Shi, J.~Si, L.~Lv, L.~Guan, B.~Hao, Z.~Tie, D.~Wang, C.~Xing, and T.~Q. Quek, ``Intelligent covert communication: Recent advances and future research trends,'' \emph{Engineering}, vol.~44, pp. 101--111, 2025. [Online]. Available: \url{https://www.sciencedirect.com/science/article/pii/S2095809924007227}
\BIBentrySTDinterwordspacing

\bibitem{6}
B.~A. Bash, D.~Goeckel, D.~Towsley, and S.~Guha, ``Hiding information in noise: fundamental limits of covert wireless communication,'' \emph{IEEE Communications Magazine}, vol.~53, no.~12, pp. 26--31, 2015.

\bibitem{22}
Y.-A. Xie, J.~Kang, D.~Niyato, N.~T.~T. Van, N.~C. Luong, Z.~Liu, and H.~Yu, ``Securing federated learning: A covert communication-based approach,'' \emph{IEEE Network}, vol.~37, no.~1, pp. 118--124, 2023.

\bibitem{23}
\BIBentryALTinterwordspacing
B.~McMahan, E.~Moore, D.~Ramage, S.~Hampson, and B.~A.~y. Arcas, ``{Communication-Efficient Learning of Deep Networks from Decentralized Data},'' in \emph{Proceedings of the 20th International Conference on Artificial Intelligence and Statistics}, ser. Proceedings of Machine Learning Research, A.~Singh and J.~Zhu, Eds., vol.~54.\hskip 1em plus 0.5em minus 0.4em\relax PMLR, 20--22 Apr 2017, pp. 1273--1282. [Online]. Available: \url{https://proceedings.mlr.press/v54/mcmahan17a.html}
\BIBentrySTDinterwordspacing

\bibitem{26}
Y.~Feng, Y.~Jiang, and Y.~Wang, ``Gan-based covert communications against an adversary with uncertain detection threshold in federated learning networks,'' in \emph{2023 International Conference on Networking and Network Applications (NaNA)}, 2023, pp. 613--618.

\bibitem{27}
Y.~Wen, Y.~Huo, J.~li, J.~Qian, and K.~Wang, ``Generative adversarial network-aided covert communication for cooperative jammers in ccrns,'' \emph{IEEE Transactions on Information Forensics and Security}, vol.~PP, pp. 1--1, 01 2025.

\bibitem{28}
I.~Goodfellow, J.~Pouget-Abadie, M.~Mirza, B.~Xu, D.~Warde-Farley, S.~Ozair, A.~Courville, and Y.~Bengio, ``Generative adversarial networks,'' \emph{Advances in Neural Information Processing Systems}, vol.~3, 06 2014.

\bibitem{31}
\BIBentryALTinterwordspacing
R.~Chinnasamy, M.~Subramanian, S.~V. Easwaramoorthy, and J.~Cho, ``Deep learning-driven methods for network-based intrusion detection systems: A systematic review,'' \emph{ICT Express}, vol.~11, no.~1, pp. 181--215, 2025. [Online]. Available: \url{https://www.sciencedirect.com/science/article/pii/S2405959525000050}
\BIBentrySTDinterwordspacing

\bibitem{13}
\BIBentryALTinterwordspacing
T.~B. Ahammed, R.~Patgiri, and S.~Nayak, ``A vision on the artificial intelligence for 6g communication,'' \emph{ICT Express}, vol.~9, no.~2, pp. 197--210, 2023. [Online]. Available: \url{https://www.sciencedirect.com/science/article/pii/S2405959522000741}
\BIBentrySTDinterwordspacing

\bibitem{32}
Z.~Zhang, Y.~Xiao, Z.~Ma, M.~Xiao, Z.~Ding, X.~Lei, G.~K. Karagiannidis, and P.~Fan, ``6g wireless networks: Vision, requirements, architecture, and key technologies,'' \emph{IEEE Vehicular Technology Magazine}, vol.~14, no.~3, pp. 28--41, 2019.

\bibitem{15}
K.~Yang, T.~Jiang, Y.~Shi, and Z.~Ding, ``Federated learning via over-the-air computation,'' \emph{IEEE Transactions on Wireless Communications}, vol.~PP, pp. 1--1, 01 2020.

\bibitem{33}
Y.~Liu, W.~Hongsheng, M.~Peng, J.~Guan, J.~Xu, and Y.~Wang, ``Deepga: A privacy-preserving data aggregation game in crowdsensing via deep reinforcement learning,'' \emph{IEEE Internet of Things Journal}, vol.~7, pp. 4113--4127, 05 2020.

\bibitem{16}
S.~Park and W.~Choi, ``On the differential privacy in federated learning based on over-the-air computation,'' \emph{IEEE Transactions on Wireless Communications}, vol.~23, no.~5, pp. 4269--4283, 2024.

\bibitem{17}
G.~Ang, R.~Xiaoyu, D.~Bin, S.~Xinshun, and Z.~Jiankang, ``Joint active and passive beamforming design in intelligent reflecting surface (irs)-assisted covert communications: A multi-agent drl approach,'' \emph{China Communications}, vol.~21, no.~9, pp. 11--26, 2024.

\bibitem{34}
Y.~Zhang, Y.~Zhang, J.~Wang, S.~Xiao, and W.~Tang, ``Distance-angle beamforming for covert communications via frequency diverse array: Toward two-dimensional covertness,'' \emph{IEEE Transactions on Wireless Communications}, vol.~22, no.~12, pp. 8559--8574, 2023.

\bibitem{18}
X.~Hou, J.~Wang, C.~Jiang, X.~Zhang, Y.~Ren, and m.~Debbah, ``Uav-enabled covert federated learning,'' \emph{IEEE Transactions on Wireless Communications}, vol.~PP, 02 2023.

\bibitem{35}
Z.~Yao, W.~Cheng, W.~Zhang, T.~Zhang, and H.~Zhang, ``The rise of uav fleet technologies for emergency wireless communications in harsh environments,'' \emph{IEEE Network}, vol.~36, no.~4, pp. 28--37, 2022.

\bibitem{19}
T.~Q. Duong, L.~D. Nguyen, B.~Narottama, J.~A. Ansere, D.~V. Huynh, and H.~Shin, ``Quantum-inspired real-time optimization for 6g networks: Opportunities, challenges, and the road ahead,'' \emph{IEEE Open Journal of the Communications Society}, vol.~3, pp. 1347--1359, 2022.

\bibitem{7}
B.~A. Bash, D.~Goeckel, and D.~Towsley, ``Limits of reliable communication with low probability of detection on awgn channels,'' \emph{IEEE Journal on Selected Areas in Communications}, vol.~31, no.~9, pp. 1921--1930, 2013.

\bibitem{36}
\BIBentryALTinterwordspacing
F.~Nielsen, ``On a generalization of the jensen–shannon divergence and the jensen–shannon centroid,'' \emph{Entropy}, vol.~22, no.~2, 2020. [Online]. Available: \url{https://www.mdpi.com/1099-4300/22/2/221}
\BIBentrySTDinterwordspacing

\bibitem{37}
I.~Gulrajani, F.~Ahmed, M.~Arjovsky, V.~Dumoulin, and A.~Courville, ``Improved training of wasserstein gans,'' in \emph{Proceedings of the 31st International Conference on Neural Information Processing Systems}, ser. NIPS'17.\hskip 1em plus 0.5em minus 0.4em\relax Red Hook, NY, USA: Curran Associates Inc., 2017, p. 5769–5779.

\bibitem{20}
H.~Poddar, S.~Ju, D.~Shakya, and T.~S. Rappaport, ``A tutorial on nyusim: Sub-terahertz and millimeter-wave channel simulator for 5g, 6g, and beyond,'' \emph{IEEE Communications Surveys and Tutorials}, vol.~26, no.~2, pp. 824--857, 2024.

\end{thebibliography}
 Generated by IEEEtran.bst, version: 1.14 (2015/08/26)

\vfill

\end{document}